\documentclass[pre]{revtex4}

\usepackage{amsmath}    
\usepackage{amsfonts}
\usepackage{amssymb}
\usepackage{graphicx}   
\usepackage{dcolumn}
\usepackage{bm}

\begin{document}

\title{Filamentation instability in a quantum plasma}

\author{A. Bret}
\email{antoineclaude.bret@uclm.es}
 \affiliation{ETSI Industriales, Universidad de Castilla-La
Mancha, 13071 Ciudad Real, Spain}

\date{\today }

\begin{abstract}
The growth rate of the filamentation instability triggered when a
diluted cold electron beam passes through a cold
plasma is evaluated using the quantum hydrodynamic equations.
Compared with a cold fluid model, quantum effects reduce both the
unstable wave vector domain and the maximum growth rate. Stabilization of large wave vector modes is always achieved,
 but significant reduction of the maximum growth rate depends on a
dimensionless parameter that is provided. Although calculations are extended to the relativistic regime, they are mostly relevant to the non-relativistic one.
\end{abstract}

\maketitle

\section{Introduction}
There is currently a growing interest in quantum plasmas in
connection with many areas of physics like small electronic
devices, astrophysics or laser plasma interaction (see Refs.
\cite{Ren2007,Haas2005} and references therein). As far as plasma
instabilities are concerned, the quantum theory of the two-stream
instability has been elaborated in Refs.
\cite{Anderson2002,Haas2000,HaasManfredi2003,HaasGarcia2003} while
quantum effects on the filamentation instability are still to
evaluate. The purpose of this work is to deal with this issue
which could be relevant for some astrophysical settings, or the
Fast Ignition Scenario for Inertial Confinement Fusion
\cite{Tabak} where a relativistic electron beam is supposed to
reach the partially degenerate dense core of a pre-compressed
target.

We thus consider an infinite and homogenous cold relativistic
electron beam of velocity $\mathbf{V}_b$, relativistic factor
$\gamma_b$ and density $n_b$ entering a cold plasma of electronic
density $n_p$. Ions form a fixed neutralizing background of
density $n_b+n_p$ and the beam prompts a return current in the
plasma with velocity $\mathbf{V}_p$ such as
$n_p\mathbf{V}_p=n_b\mathbf{V}_b$. In a first attempt to quantity
quantum effects for the filamentation instability, we implement
the quantum fluid formalism relying on the conservation equations
for the beam ($j=b$) and the plasma ($j=p$),
\begin{equation}\label{eq:conser}
   \frac{\partial n_j}{\partial t}+\nabla\cdot(n_j \mathbf{v}_j)=0
\end{equation}
and the force equation with a Bohm potential term \cite{Haas2005},
\begin{equation}\label{eq:force}
   \frac{\partial \mathbf{p}_j}{\partial
   t}+(\mathbf{v}_j\cdot\nabla)\mathbf{p}_j=-q\left(\mathbf{E} + \frac{\mathbf{v}_j\times
   \mathbf{B}}{c}\right)+\frac{\hbar^2}{2
   m}\nabla\left(\frac{\nabla^2\sqrt{n_j}}{\sqrt{n_j}}\right),
\end{equation}
where $q>0$ and $m$ are the charge and mass of the electron, $n_j$
the density of species $j$, and $p_j$ its momentum. Quantum corrections are clearly contained within the Bohm potential, which can be obtained from the moments of the non-relativistic Wigner function \cite{Haas2005}. The present calculations are here extended to the relativistic regime, but a correct treatment of quantum effects in that limit should rely on the moments of a \emph{relativistic} Wigner function  such as the one described in Refs. \cite{Bialynicki1991,Shin1996}. Nevertheless the modified force equation (\ref{eq:force}) has to our knowledge no simple relativistic counterpart yet, and present relativistic results make sense physically. This is why we found relevant to deal with the relativistic regime with the present equations. Although the following theory is only correct in the non-relativistic case,  it could also be interesting to compare future relativistic results with the present ones.

Assuming
quantities perturbed according to  $\exp(i \mathbf{k}\cdot
\mathbf{r} - i\omega t)$, we find the linearized equations,
\begin{equation}\label{eq:conserline}
    n_{j1}=n_{j0}\frac{\mathbf{k}\cdot \mathbf{v}_{j1}}{\omega-\mathbf{k}\cdot
    \mathbf{v}_{j0}},
\end{equation}
and
\begin{equation}\label{eq:forceL}
 i m \gamma_j (\mathbf{k}\cdot
\mathbf{v}_{j0}-\omega)\left(\mathbf{v}_{j1}+\frac{\gamma_j^2}{c^2}(\mathbf{v}_{j0}\cdot
\mathbf{v}_{j1})\mathbf{v}_{j0}\right)
  = -q\left(\mathbf{E}_{1}+\frac{\mathbf{v}_{j0}\times
   \mathbf{B}_1}{c}\right)-i\frac{\hbar
   k^2}{4m}\frac{n_{j1}}{n_{j0}}\mathbf{k},
\end{equation}
where subscript 0 and 1 stand for the equilibrium and perturbed
quantities, and $\gamma_j$ for the relativistic factor of the
unperturbed beam and return current. Note that the diluted beam
hypothesis $n_b\ll n_p$ implies a non-relativistic return current
with $\gamma_p=1$.

We now choose to align the beam velocity along the $z$ axis and
$\mathbf{k}$ along the $x$ axis, and calculate the dispersion
equation following a quite standard procedure
\cite{BretPoPFluide}. Expressing the current $\mathbf{J}$ in terms
of the electric field $\mathbf{E}$ and inserting it into a
combination of Maxwell's equations,
\begin{equation}\label{eq:Maxwell}
  \frac{c^2}{\omega^2}\mathbf{k}\times(\mathbf{k}\times \mathbf{E})+\mathbf{E} + \frac{4
  i
  \pi}{\omega}\mathbf{J} = 0,
\end{equation}
allows for the derivation of the dielectric tensor. Its
determinant finally gives the dispersion equation
\begin{eqnarray}\label{eq:disper}
 \frac{x^4 Z^2\alpha^2(\gamma_b-1)^2} {(x^2 - \Theta Z^4)^2(x^2\gamma_b-\Theta Z^4)^2}
 &=&
 \left[ 1 + \frac{1}{\Theta Z^4-x^2} +  \frac{\alpha }{\Theta  Z^4 - x^2\gamma_b}
 \right]\\
 &\times&\left[x^2-1-\frac{\alpha}{\gamma_b^3}-\frac{Z^2}{\beta^2}
+\frac{\alpha^2Z^2}{\Theta Z^4-x^2 } + \frac{\alpha Z^2}{\Theta
Z^4 - x^2\gamma_b}
     \right],\nonumber
\end{eqnarray}
in terms of,
\begin{equation}\label{eq:param}
    x=\frac{\omega}{\omega_p},~Z=\frac{kV_b}{\omega_p},~\beta=\frac{V_b}{c},~\alpha=\frac{n_b}{n_p},
\end{equation}
where $\omega_p$ is the electronic plasmas frequency, and
\begin{equation}\label{eq:theta}
    \Theta=\left(\frac{\hbar\omega_p}{2mV_b^2}\right)^2 \equiv  \frac{\Theta_c}{\beta^4}
\end{equation}
with $\Theta_c = (\hbar\omega_p/2mc^2)^2$. Although this quantity
remains  small even in a very dense plasma with
$\Theta=1,3.10^{-7}$ assuming $n_p=10^{26}$ cm$^{-3}$ and $V_b\sim
c$, we will check in the sequel that quantum effects may not be
negligible for the filamentation instability.

\section{Quantum effects analysis}
\begin{figure}[t]
\begin{center}
\includegraphics[width=0.45\textwidth]{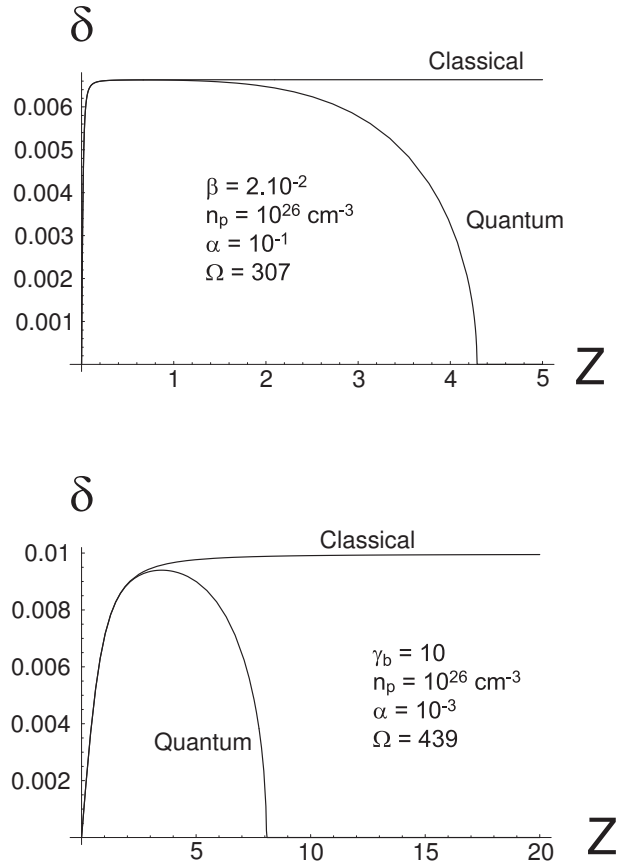}
\end{center}
\caption{Growth rate $\delta$ in $\omega_p$ units of the
filamentation instability for a diluted non-relativistic (upper graph) and relativistic (lower graph) beam. The plasma density in both cases is
$n_p=10^{26}$ cm$^{-3}$ yielding $\Theta_c=1,3.10^{-7}$. The range
of unstable wave vectors is dramatically reduced whereas the most
unstable mode and its growth rate are only weakly affected.
Parameter $\Omega$ given by Eq. (\ref{eq:parameter}) is here 307 and 439 respectively.} \label{fig:1}
\end{figure}

Let us start plotting  on Figure \ref{fig:1} the growth rate in
terms of $Z$ for a diluted  beam passing through a
dense plasma,in the non-relativistic and relativistic regimes. We make the comparison between the calculations for
$\Theta_c=0$ and $\Theta_c=1,3.10^{-7}$. In both cases, it is straightforward that
although the maximum growth rate is not modified (and we will see
why later), quantum effects introduce a cut off at large $Z$
whereas the growth rate just saturates in the classical limit
$\Theta=0$.

The cut off value of $Z$ can be calculated exactly in terms of
$\Theta$ noting that a vanishing growth rate implies that Eq.
(\ref{eq:disper}) be verified with $x=0$ since the real part of
the root is 0 for the filamentation instability. The resulting
expression can be solved exactly in terms of $Z$ and gives the cut
off wave vector $Z_m$,
\begin{equation}\label{eq:ZmExact}
    Z_m^2 = \frac{\beta^2( \alpha  + \gamma_b^3)
    \left( {\sqrt{1 + \frac{4\alpha ( 1 + \alpha)
             \gamma_b^6}{\Theta{\beta }^2
             ( \alpha  + \gamma_b^3)^2}}}
      -1\right)}{2\gamma_b^3}.
\end{equation}
Accounting for $\alpha\ll 1$, we find
\begin{equation}\label{eq:ZmApprox}
    Z_m^2 \sim  \frac{\beta^2 }{2}\left( {\sqrt{1 + \frac{4\alpha }{\Theta\beta^2 }}}-1\right).
\end{equation}
It turns out that the parameters $4\alpha/\Theta\beta^2$ measures
the strength of quantum effects as far as $Z_m$ is concerned. For
the sake of physical interpretation, let us consider the most
common case where this parameter is large with respect to 1, and
set $\beta\sim 1$. Equation (\ref{eq:ZmApprox}) can be developed
and yields
\begin{equation}\label{eq:CutOffk}
 k_m \sim \sqrt{\frac{2m\omega_b}{\hbar}},
\end{equation}
which is the inverse of the beam quantum wavelength. Simply put,
perturbations which scale length is smaller than the beam quantum
wavelength are relaxed and do not grow. Also remarkable, the
$\gamma_b$ dependance is lost between Eqs. (\ref{eq:ZmExact}) and
(\ref{eq:ZmApprox}) because of the diluted beam hypothesis, so
that it can be said that to a large extent, quantum corrections to
the instability domain in the relativistic regime do not depend on
the beam energy as long as they are described with the correction term inserted in Eq. (\ref{eq:force}).

\begin{figure}[t]
\begin{center}
\includegraphics[width=0.45\textwidth]{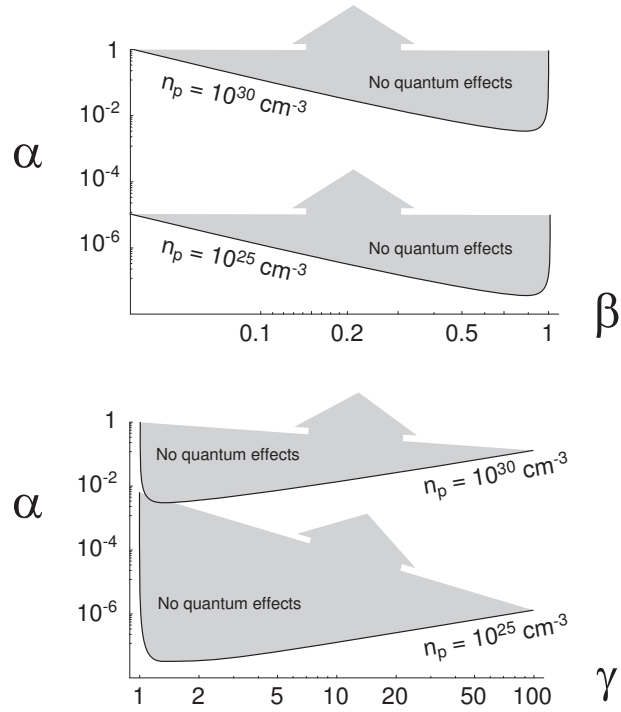}
\end{center}
\caption{Region of the ($\alpha,\beta$ or $\gamma_b$) plan where
$\Omega \gg 1$ for $n_p=10^{25}$ and $10^{30}$ cm$^{-3}$. The
curves picture the border $\Omega=1$ and the shaded areas
indicates the side of the border where quantum effects are
negligible with $\Omega\gg 1.$} \label{fig:0}
\end{figure}

While quantum effects on $Z_m$ are necessarily important since
$Z_m=\infty$ in the classical case, effects on the largest growth
rate are even more relevant for the physics of the system. If we
recall that in the classical limit, the growth rate $\delta$
saturates for $Z \gtrsim \beta$ with $\delta \sim
\beta\sqrt{\alpha/\gamma_b}$ \cite{BretPRE2004}, the occurrences
of $\Theta$ in dispersion equation (\ref{eq:disper}) indicate that
quantum effects near the most unstable wave vectors are negligible
if
\begin{eqnarray}\label{eq:Condition12}
    \Theta Z^4|_{Z\sim\beta} &\ll&
    x^2|_{x\sim\beta\sqrt{\alpha/\gamma_b}}, \nonumber\\
       \Theta Z^4|_{Z\sim\beta} &\ll& \gamma_bx^2|_{x\sim\beta\sqrt{\alpha/\gamma_b}}.
\end{eqnarray}
These two conditions are clearly met if the first one is
satisfied,
\begin{equation}\label{eq:Condition}
    \Theta \beta^4  \ll  \beta^2\alpha/\gamma_b \Leftrightarrow \frac{\alpha}{\Theta\gamma_b\beta^2}\gg 1.
\end{equation}
We here recover the $\alpha/\Theta\beta^2$ parameter already
evidenced when dealing with the largest unstable wave vector, but
we find it must dominate the beam relativistic factor if quantum
effects are to be negligible with respect to the most unstable
mode. Therefore, depending on the inequality (\ref{eq:Condition}),
we will find situations where quantum effects stabilize the
instability at large $Z$ while changing, or not, the most unstable
mode. The parameter
\begin{equation}\label{eq:parameter}
   \Omega=\frac{\alpha}{\Theta\gamma_b\beta^2} = \frac{\alpha\beta^2}{\Theta_c\gamma_b},
\end{equation}
is thus the most relevant one from the physical point of view
because a change in the fastest growing mode brings more physical
consequences than the stabilization of the non-dominant large $Z$
modes. $\Omega \gg 1$ implies few quantum effects for the most
unstable mode whereas $\Omega \ll 1$ means strong quantum effects.

The physical interpretation of $\Omega$ can be elaborated after
the one we used for $Z_m$ in Eqs.
(\ref{eq:ZmApprox},\ref{eq:CutOffk}). We found that wavelengths
larger than the beam quantum wavelength are stabilized. Let us now
construct the ratio of $k_m$ as given by Eq. (\ref{eq:CutOffk})
with the plasma skin depth $\omega_p/c$ wave number, where the
instability reaches its maximum in the classical regime. We take
the fourth power of this quantity and consider the relativistic
beam regime and find,
\begin{equation}\label{eq:OmegaPhys}
    \left(\frac{k_m}{\omega_p/c}\right)^4=\frac{n_b}{n_p}\left(\frac{2mc^2}{\hbar\omega_p}\right)^2=\frac{\alpha}{\Theta_c}.
\end{equation}
We thus check that once added a relativistic contraction factor
$1/\gamma_b$, the parameter $\Omega$ can be constructed from the
ratio of the plasma skin depth $\lambda_s$ to the beam quantum
wavelength $k_m^{-1}$. When $\lambda_s$ is much smaller than
$k_m^{-1}$, the perturbations which should give rise to the most
unstable modes are ``blurred'' by quantum effects, resulting in a
tamed instability.

In the non-relativistic regime, $\Omega=\alpha\beta^2/\Theta_c$
 can be traced back, after some manipulations exploiting the
identity $n_b/n_p=\omega_b^2/\omega_p^2=V_p/V_b$, to
\begin{equation}\label{eq:OmegaPhysNR}
\Omega = 4\left(\frac{c/\omega_p}{\lambda_e}\right)^2,
\end{equation}
where $\lambda_e=\hbar/mV_b$ is the de Broglie length of a beam
electron. Quantum effects are here important for the maximum
growth rate when the beam electron de Broglie length is much
larger than the plasma skin depth.

We picture on Figure \ref{fig:0} the region of the
($\alpha,\beta$ or $\gamma_b$) plan where $\Omega \gg 1$ for two
different plasma densities. From Eq. (\ref{eq:parameter}) and Fig.
\ref{fig:0}, we infer that quantum effects tend to be important
for beams with $\beta\ll 1$ or $\gamma_b\gg 1$, that is, the
non-relativistic and ultra-relativistic limits. The smallest beam
densities are required in the intermediate regime for
$\gamma_b=\sqrt{2}$ with $\alpha\ll 2\Theta_c$.

\begin{figure}[t]
\begin{center}
\includegraphics[width=0.45\textwidth]{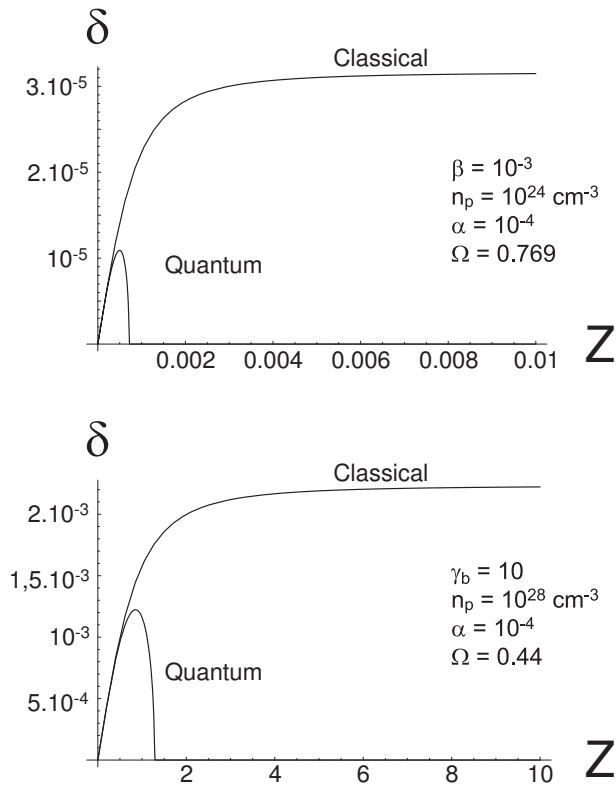}
\end{center}
\caption{Growth rate $\delta$ in $\omega_p$ units of the
filamentation instability for a non-relativistic (upper graph) and
a relativistic beam (lower graph) and $\Omega \ll 1$. Here the
most unstable mode and its growth rate are significantly modified
by quantum effects.} \label{fig:2}
\end{figure}

Figure \ref{fig:1}  already displays two systems with $\Omega=307$ and 439 $\gg
1$. As expected, according to the previous analysis, the maximum
growth rates are weakly affected by quantum effects whereas the
ranges of the instability are reduced. Figure \ref{fig:2} pictures
now one non-relativistic and one relativistic situations with
small $\Omega$ parameters. The maximum growth rate is
significantly reduced in both cases so that quantum effects should
have easily observable consequences. Noteworthily, the
relativistic scenario requires a much higher plasma density than
the non-relativistic one, where $\Omega$ can be lowered down
through the $\beta^2$ factor at the numerator. Quantum effects in
the relativistic regime should thus be expected mostly in astrophysical
settings, as already noted in the context of quantum
magnetohydrodynamics \cite{Haas2005} for example.

\section{Discussion and Conclusion}
We have implemented the quantum fluid equations to evaluate
quantum corrections to the filamentation instability. Quantum
effects stabilize the instability for large wave vectors and can
significantly reduce the most unstable mode if parameters $\Omega$
defined by Eq. (\ref{eq:parameter}) is much smaller than unity.
Under such conditions, quantum effect can reduce the filamentation
instability. We could check that there is eventually no quantum
cancelation of the instability. Indeed, a numerical exploration
of the regime $\Omega\ll 1$ shows that the maximum growth rate
behaves like $\sqrt{\Omega}$ in this limit. Given the requirements
on the parameters for quantum corrections to become significant,
the necessity of a quantum treatment may be restricted to some
astrophysical scenarios when relativistic beams are involved. Even
in the conditions of the Fast Ignition Scenario for Inertial
Confinement Fusion, cold fluid quantum corrections are only about
$10\%$, as $n_p=10^{26}$ cm$^{-3}$, $n_b=10^{21}$ cm$^{-3}$ and
$\gamma_b=5$ yields $\Omega=159$. Nevertheless, a correct relativistic treatment should rely on the moments of the relativistic Wigner function, so that present results are only valid in the non-relativistic case while their relevance for the relativistic regime may be only qualitative.

Let us add that it would be very interesting to
investigate the joined effects of temperature and quantum
degeneracy. In the classical relativistic limit, it is well known
that kinetic effects can literally suppress filamentation
\cite{Silva2002,BretPoPFilaWeibel}. Since quantum effects can also
have some stabilizing function,  a quantum kinetic treatment of
the problem could well unravel even more pronounced stabilizing
quantum effects than those investigated here.

In the same way, it has been known for long that filamentation instability can be significantly reduced, if not completely suppressed, in a magnetized plasma \cite{Cary1981}. While various quantum effects have been recently investigated in the case of a magnetized plasma \cite{ShuklaPhysLettA2006,ShuklaJPP2006,ShuklaPoP2006}, the behavior of the filamentation instability in such settings should be a subject of much interest.

\section{Acknowledgements}
This work has been  achieved under projects FTN 2003-00721 of the
Spanish Ministerio de Educaci\'{o}n y Ciencia and PAI-05-045 of
the Consejer\'{i}a de Educaci\'{o}n y Ciencia de la Junta de
Comunidades de Castilla-La Mancha.


\end{document}